\documentclass[twocolumn]{aastex62}

\newcommand{\teff}{T_{\rm{eff}}}
\usepackage{xcolor}

\begin{document}

\title{Visual Orbits of Spectroscopic Binaries with the CHARA Array. I. HD 224355}

\author{Kathryn V. Lester}
\affil{Center for High Angular Resolution Astronomy and Department of Physics \& Astronomy, \\ Georgia State University, Atlanta, GA 30302 USA}
\email{lester@astro.gsu.edu}

\author{Douglas R. Gies}
\affil{Center for High Angular Resolution Astronomy and Department of Physics \& Astronomy, \\ Georgia State University, Atlanta, GA 30302 USA}

\author{Gail H. Schaefer} 
\affil{The CHARA Array, Mount Wilson Observatory, Mount Wilson, CA 91023 USA}

\author{Christopher D. Farrington}
\affil{The CHARA Array, Mount Wilson Observatory, Mount Wilson, CA 91023 USA}

\author{John D. Monnier}  
\affil{Department of Astronomy, University of Michigan, Ann Arbor, MI 48109 USA}

\author{Theo ten Brummelaar}  
\affil{The CHARA Array, Mount Wilson Observatory, Mount Wilson, CA 91023 USA}

\author{Judit Sturmann}  
\affil{The CHARA Array, Mount Wilson Observatory, Mount Wilson, CA 91023 USA}

\author{Norman Vargas}  
\affil{The CHARA Array, Mount Wilson Observatory, Mount Wilson, CA 91023 USA}

\begin{abstract}

We present the visual orbit of the double-lined spectroscopic binary HD 224355 from interferometric observations with the CHARA Array, as well as an updated spectroscopic analysis using echelle spectra from the Apache Point Observatory 3.5m telescope.  By combining the visual and spectroscopic orbital solutions, we find the binary components to have masses of $M_1 = 1.626 \pm 0.005 M_\odot$ and $M_2 = 1.608 \pm 0.005 M_\odot$, and a distance of $d = 63.98 \pm 0.26$ pc.  Using the distance and the component angular diameters found by fitting spectrophotometry from the literature to spectral energy distribution models, we estimate the stellar radii to be $R_1 =  2.65\pm0.21 R_\odot$ and $R_2 =  2.47\pm0.23 R_\odot$.  We then compare these observed fundamental parameters to the predictions of stellar evolution models, finding that both components are evolved towards the end of the main sequence with an estimated age of $1.9$~Gyr.

\end{abstract}

\section{Introduction}
Accurate fundamental parameters of binary stars have become important tools for testing models of stellar evolution and interiors. Systems with uncertainties in stellar mass and radius of less than 3\% are used for calibrating the physics within evolutionary models \citep{ct16, ct18} and creating mass-radius and mass-luminosity relationships for use with single stars \citep[e.g.,][]{torres10, eker15, moya18}. These models and relationships are then used in other areas of stellar astronomy, such as calibrating asteroseismic scaling relations \citep{astero} and determining the properties of exoplanet host stars and their exoplanets \citep{enoch10}.

Binary systems with very precise parameters are often eclipsing, double-lined systems whose radial velocities and light curves are used to determine the component masses and radii. However, most eclipsing binaries have short orbital periods due to the higher probability of occultation in systems with a separation not much larger than the sum of the radii. For example, 82\% of the stars in the \citet{torres10} sample have orbital periods less than 7 days. These short orbital periods and small separations can introduce several challenges -- such as the presence of a distant tertiary companion, reflection effects, and tidal distortions and locking -- that can alter the stellar interiors, atmospheric and observational properties, and evolutionary paths \citep{hurley02, tokovinin06}. 

Therefore, the stars in close binary systems may not evolve like single stars or be the best test subjects for stellar evolution models. We need to expand studies to longer period, non-interacting double-lined spectroscopic binary (SB2) systems in order to look for systematic differences between the parameters of short and longer period binaries. Even though longer period SB2 systems are less likely to be eclipsing, their fundamental parameters can be determined by resolving the orbital motion in the plane of the sky. This visual orbit allows for the determination of several orbital parameters, such as inclination and angular semi-major axis, and provides masses and distances when combined with the spectroscopic orbit. Long baseline optical interferometers can resolve the relative motion of the secondary component around the primary on milliarcsecond (mas) scales \citep[e.g.,][]{hummel93, boden99, deepak09}, which opens up dozens of nearby SB2 systems as candidates for measuring visual orbits and determination of their fundamental parameters \citep{halbwachs81}. 

For this purpose, we began an observing campaign with the CHARA Array interferometer to measure the visual orbits of 11 nearby SB2 systems with component stars of B, A and F spectral types. One binary in our sample is HD~224355\footnote{HR 9059, HIP 118077, V1022 Cas; \ $\alpha$ = 23 57 08.47,  $\delta$ = +55 42 20.53 (J2000); \ $V=5.6$ mag}, which was discovered to be a double-lined binary by \citet{plaskett20}. Spectroscopic orbits of HD 224355 were completed by \citet{harper23},  \citet{imbert77}, and most recently by \citet{fekel10}, who obtained over a hundred observations using three echelle spectrographs to determine precisely the orbital parameters and minimum masses of this system. While \citet{otero06} noted a partial primary eclipse in \textit{Hipparcos} photometry \citep{hipparcos1}, a secondary eclipse was not observed due to gaps in coverage at the predicted phase. 

We present a visual orbit of HD 224355 using observations from the CHARA Array, as well as an updated spectroscopic analysis using echelle spectra from the Apache Point Observatory, in order to determine the fundamental parameters of this system. Section~\ref{section:spec} describes our spectroscopic observations and radial velocity analysis, while Section~\ref{section:inter} describes our interferometric observations and analysis.  We describe the individual and combined methods of fitting for orbital parameters in Section~\ref{section:orbit} and present the derived stellar parameters in Section~\ref{section:param}.

\section{Spectroscopy}\label{section:spec}

\subsection{ARCES Observations}
We obtained 16 nights of data using the Astrophysical Research Consortium echelle spectrograph \citep[ARCES,][]{arces} on the Apache Point Observatory (APO) 3.5m telescope between 2015 December - 2018 June. ARCES covers $\lambda 3500-10500$\AA\  over 107 orders at an average resolving power of $R\sim30,000$.  Data were reduced using standard IRAF procedures, including bias subtraction, cosmic ray removal, one-dimensional flat fielding, and wavelength calibration from Thorium-Argon lamp exposures. All spectra were corrected to the heliocentric frame and transformed onto a standard logarithmic wavelength grid.  The echelle blaze function was removed using the procedure of \citet{kolbas15}, where templates for the blaze function were created from polynomial fits to orders free of strong absorption lines. These templates were interpolated to the orders where strong absorption lines were present, such as the H$\alpha$ order. Normalized spectra for each echelle order were then created by dividing the observed spectra by the blaze templates.

\subsection{Radial Velocities}
We measured the radial velocities ($V_r$) of our ARCES spectra using the TwO-Dimensional CORrelation (TODCOR) procedure of \citet{todcor}, which computes the correlation coefficient between the observed spectrum and a template composite spectrum across a grid of primary and secondary radial velocities. Templates were taken from BLUERED\footnote{\href{http://www.inaoep.mx/~modelos/bluered/bluered.html}{http://www.inaoep.mx/$\sim$modelos/bluered/bluered.html}} model spectra \citep{bluered} using the atmospheric parameters estimated by \citet{fekel10} ($T_{\rm{eff}_1}  =  6300$ K, $T_{\rm{eff}_2} = 6300$ K; $\log g_1 = 4.0, \log g_2 = 4.0$; $V_1 \sin i = 11.5,  V_2 \sin i =9.0$ km~s$^{-1}$; $f_2/f_1 = 0.9$) and solar metallicity.

We ran TODCOR individually for each echelle order in the range $4500-6600$\AA. Because the primary and secondary components have very similar template spectra, we manually identified and corrected any orders where the component velocities were switched. We then computed the final radial velocities for each night from the weighted average of the velocities from each echelle order and the uncertainties from the standard deviation in all orders. Our results are listed in Table~\ref{rvtable}, along with the residuals to the combined solution found in Section~\ref{vbsbfit}. TODCOR also estimates the flux ratio for each echelle order, all with similar results.  For example, the fitted flux ratio for the H$\alpha$ order is $f_2/f_1 = 0.95 \pm 0.06$.		

\begin{deluxetable*}{ccRRRRRR}
\tablewidth{0pt}
\tablecaption{  Radial Velocity Measurements for HD 224355\label{rvtable}  }
\tablehead{ \colhead{UT Date} & \colhead{Orbital} & \colhead{$V_{r1}$} & \colhead{$\sigma_1$} & \colhead{Residual}  & \colhead{$V_{r2}$} & \colhead{$\sigma_2$} & \colhead{Residual}   \\  
\colhead{(HJD-2,400,000)} & \colhead{Phase} & \colhead{(km~s$^{-1}$)} & \colhead{(km~s$^{-1}$)} & \colhead{(km~s$^{-1}$)} & \colhead{(km~s$^{-1}$)} & \colhead{(km~s$^{-1}$)} & \colhead{(km~s$^{-1}$)}}
\startdata 	
57357.6250  &  0.25  &  $ -35.50$  	&  0.28   &  $ -0.07$  &  $  59.53$  &   0.28	&  $  0.09$  \\ 
57413.5664  &  0.85  &  $  69.37$  	&  0.22   &  $ -1.02$  &  $ -48.12$  &   0.32	&  $ -0.56$  \\ 
57645.7656  &  0.95  &  $ 101.60$  	&  0.26   &  $  0.48$  &  $ -77.96$  &   0.29	&  $  0.67$  \\ 
57682.7109  &  0.99  &  $  93.78$  	&  0.31   &  $ -0.08$  &  $ -72.21$  &   1.21	&  $ -0.93$  \\ 
57676.5938  &  0.49  &  $ -30.95$  	&  0.28   &  $ -0.35$  &  $  54.11$  &   1.29 	&  $ -0.45$  \\ 
57708.6211  &  0.12  &  $   6.10$  	&  0.46   &  $  0.09$  &  $  16.91$  &   1.07 	&  $ -0.63$  \\ 
57711.6289  &  0.37  &  $ -40.26$  	&  0.37   &  $  0.27$  &  $  64.62$  &   0.21 	&  $  0.02$  \\ 
57737.5820  &  0.50  &  $ -27.67$  	&  0.38   &  $  0.62$  &  $  52.37$  &   0.15 	&  $  0.15$  \\ 
57759.5664  &  0.31  &  $ -40.18$  	&  0.42   &  $  0.55$  &  $  64.85$  &   0.14 	&  $  0.06$  \\ 
57998.7461  &  0.99  &  $  94.29$  	&  0.23   &  $ -0.36$  &  $ -72.81$  &   0.86	&  $ -0.73$  \\ 
58027.6406  &  0.36  &  $ -41.16$  	&  0.51   &  $ -0.50$  &  $  64.08$  &   0.29	&  $ -0.65$  \\ 
58089.8047  &  0.48  &  $ -33.79$  	&  1.27   &  $ -2.16$  &  $  53.10$  &   1.32	&  $ -2.50$  \\ 
58114.5664  &  0.51  &  $ -27.00$  	&  0.34   &  $ -0.42$  &  $  49.68$  &   1.34	&  $ -0.82$  \\ 
58122.5977  &  0.18  &  $ -18.52$  	&  0.35   &  $  0.31$  &  $  43.31$  &   0.30 	&  $  0.66$  \\ 
58271.9336  &  0.46  &  $ -34.50$  	&  0.84   &  $ -0.82$  &  $  56.54$  &   0.16	&  $ -1.13$  \\ 
58294.9453  &  0.35  &  $ -39.36$  	&  0.23   &  $  1.57$  &  $  65.99$  &   1.42 	&  $  1.00$  \\ 
\enddata
\end{deluxetable*}

\begin{deluxetable*}{lcclccR}	
\tablewidth{0pt}
\tablecaption{CHARA/CLIMB Observing Log for HD 224355  \label{charalog}}
\tablehead{ \colhead{UT Date} & \colhead{HJD-2,400,000} & \colhead{Telescope} & \colhead{Calibrators} & \colhead{Number}  & \colhead{Number} & \colhead{$r_0$}  \\
\colhead{ } & \colhead{ } & \colhead{Configuration} & \colhead{ } & \colhead{of $V^2$}  & \colhead{of CP} & \colhead{(cm)}   }
\startdata 	
2014 Oct 05   	&  56935.7897  &  S1-W1-E1   	& HD 3360		 	  &     6	&   2	  &   13.2  \\  
2016 Sep 18  	&  57649.8375  &  S1-E1-W1	& HD 3360			  &	12	&   4	  &   8.3  	\\  
2017 Jul 02    	&  57936.9442  &  S1-E1-W1	& HD 222618, HD 222932	  &	18	&   6	  &   9.2  	\\  
2017 Jul 20   	&  57954.9226  &  S1-E1-W1	& HD 222618, HD 222932	  &	15	&   5	  &   6.2  	\\  
2017 Aug 04  	&  57970.0145  &  S1-E1-W1	& HD 222618, HD 222932	  &	 6	&   2	  &   10.0  \\   
2017 Aug 05  	&  57970.9754  &  S1-E1-W1	& HD 222618, HD 222932	  &	18	&   6	  &   8.3  	\\  
2017 Sep 07  	&  58003.7752  &  E1-W1-W2	& HD 222618, HD 222932	  &	12	&   4	  &   10.6  \\  
2017 Sep 08  	&  58004.8072  &  S1-E1-W1	& HD 222618, HD 222932	  & 	15	&   5	  &   10.2  \\  
2017 Oct 11   	&  58037.7835  &  S1-E1-W1	& HD 222618, HD 222932	  &	 9	&   3	  &   10.6  \\  
\enddata
\end{deluxetable*}

\begin{figure*}	
\centering
\epsscale{1.2}
\plotone{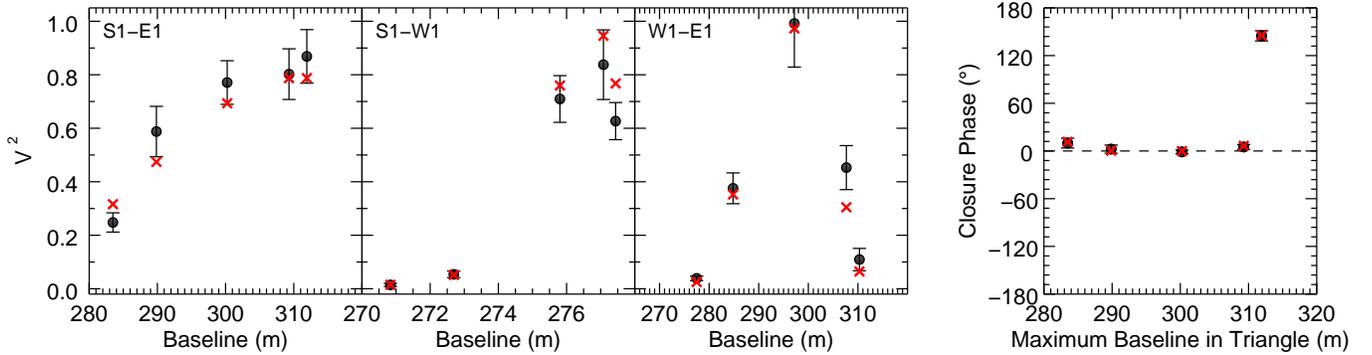}
\caption{Example squared visibilities (left) and closure phases (right) as a function of projected baseline for HD~224355 from 2017 Sept 08. The black circles represent the observed values and the red crosses represent the best-fit binary model.\label{vis}} 
\vspace{0.5cm}
\end{figure*}

 \begin{deluxetable*}{lccRRRRRc}
\tablewidth{0pt}
\tablecaption{Relative Positions for HD 224355 \label{relpos}}
\tablehead{ \colhead{UT Date} & \colhead{HJD-2,400,000} & \colhead{Orbital} & \colhead{$\rho$} & \colhead{$\theta$} &  \colhead{$\sigma_{maj}$}  &  \colhead{$\sigma_{min}$}  &  \colhead{$\phi$}  &  \colhead{$f_2/f_1$}  \\
\colhead{ } & \colhead{ } & \colhead{Phase} & \colhead{(mas)} & \colhead{(deg)} &  \colhead{(mas)}  &  \colhead{(mas)}  &  \colhead{(deg)}  &  \colhead{ }  }
\startdata 	
2014 Oct 05  	&  56935.7897  &  0.54  &  2.298 &  212.9 &   0.053 &   0.023 &    49.0 &   $0.968\pm0.016$  \\
2016 Sep 18 	&  57649.8375  &  0.28  &  2.693 &  220.9 &   0.033 &   0.023 &    13.4 &   $0.983\pm0.012$  \\
2017 Jul 02  	&  57936.9442  &  0.90  &  1.654 &    43.0 &   0.070 &   0.033 &    64.9 &   $1.000\pm0.061$  \\
2017 Jul 20 	&  57954.9226  &  0.38  &  2.940 &  218.4 &   0.021 &   0.010 &      5.8 &   $0.938\pm0.007$  \\
2017 Aug 04 	&  57970.0145  &  0.62  &  1.568 &  205.1 &   0.075 &   0.042 &  141.8 &    \nodata   		\\
2017 Aug 05 	&  57970.9754  &  0.70  &  0.677 &  188.9 &   0.041 &   0.031 &  132.2 &   $0.983\pm0.014$  \\
2017 Sep 07 	&  58003.7752  &  0.40  &  2.776 &  219.6 &   0.210 &   0.044 &    12.0 &   $0.918\pm0.010$  \\
2017 Sep 08 	&  58004.8072  &  0.49  &  2.693 &  215.1 &   0.025 &   0.015 &    41.2 &   $0.910\pm0.008$  \\
2017 Oct 11 	&  58037.7835  &  0.20  &  1.936 &  224.7 &   0.049 &   0.027 &  121.7 &   $0.871\pm0.026$  \\
\enddata  
\end{deluxetable*}

\section{Interferometry}\label{section:inter}

\subsection{CLIMB Observations}  
Interferometric observations were conducted at the CHARA Array on 9 nights between 2014 October - 2017 October. CHARA sends the light from six 1m telescopes arranged in a Y-shape with separations ranging from 34 - 330 m to one of several beam combiners operating in the optical and near infrared \citep{chara}. We used the CLassic Interferometry with Multiple Baselines beam combiner \citep[CLIMB;][]{climb}, which combines near-IR light from three telescopes in order to measure fringe visibilities and closure phases.  Our observations are listed in Table~\ref{charalog}, with the calendar and Julian dates, the telescope combination and calibrator stars used, the number of visibilities and closure phases measured, and the average Fried parameter ($r_0$) for each night. All of our observations were taken in the $K'$-band at $2.13 \mu$m, except on 2017 Oct 11 which were taken in $H$-band at $1.67 \mu$m. 

The CLIMB data were reduced with the pipeline developed by J. D. Monnier, using the general method described in \citet{monnier11} and extended to three beams \citep[e.g., ][]{kluska18}. For each observation, squared visibilities ($V^2$) were measured for each projected baseline and closure phases (CP) were measured for each closed triangle.  Calibrator stars were observed before and after the science target to complete one observation ``bracket". The $K'$-band uniform disk angular diameters from SearchCal\footnote{\href{http://www.jmmc.fr/searchcal}{http://www.jmmc.fr/searchcal}} \citep{searchcal} are $0.295\pm0.031$ mas for HD 3360, $0.668\pm0.065$ mas for HD 222618, and $0.653\pm0.017$ mas for HD 222932. In order to account for the loss of visibility from atmospheric and instrumental effects, we calculated the ratio between the observed and predicted calibrator visibilities, then divided the observed science visibilities by this factor.

\subsection{Binary Positions}
The squared visibility ($V^2$) of an interference fringe of a binary system depends on the properties of the individual components as well as the binary separation \citep{boden00},  $$ V^2_{binary} = \frac{ V^2_1 + \frac{f_2}{f_1} V^2_2 +  2\frac{f_2}{f_1} |V_1| |V_2|  \cos[2\pi(u\Delta\alpha + v\Delta\delta)] }{\big(1 + \frac{f_2}{f_1}\big)^2}$$  where $V_1$ and $V_2$ are the limb-darkened visibilities of the primary and secondary components,  $\Delta\alpha$ and $\Delta\delta$ are the relative separations in right ascension and declination in radians,  $u$ and $v$ are the spatial frequencies of the baselines projected onto the sky in radians$^{-1}$, and $f_2/f_1$ is the flux ratio. The observed visibilities therefore change over the course of one night as the projected baselines change and  throughout the orbital period as the relative positions of the components change.

\begin{deluxetable*}{lccc}
\tablewidth{0pt}
\tablecaption{Orbital Parameters for HD 224355 \label{orbpar}}
\tablehead{ \colhead{Parameter} & \colhead{SB2 only} & \colhead{VB only}  & \colhead{VB + SB2} }
\startdata		
$P$ (days) 	        		& $12.156165\pm0.000012$ &  $12.156165\tablenotemark{*}$   & $12.156160 \pm 0.000015$\\
$T$ (HJD-2400000) 		& $53282.3194\pm0.0017$   &  $53282.3194\tablenotemark{*}$ & $53282.3198 \pm 0.0017$\\
$e$                    		& $0.3117 \pm 0.0003$	&  $0.3117\tablenotemark{*}$	& $0.3117\pm 0.0003$ \\
$\omega_1$ (deg)    		& $34.45 \pm 0.06$     	&  $34.45\tablenotemark{*}$ 	& $34.46  \pm 0.05$ \\
$i$ (deg)              		&  \nodata    	 	   	&  $97.1 \pm 0.3$      		& $97.1 \pm 0.5$ \\
$a$ (mas)				&  \nodata   		   	&  $2.390 \pm 0.010$  		& $2.392 \pm 0.009$ \\
$\Omega$ (deg)         	& \nodata     		   	&  $219.4 \pm 0.2$    		& $219.4	\pm 0.2$ \\   
$\gamma$ (km~s$^{-1}$)  &$11.74 \pm 0.02 $     	&  \nodata  		        		& $11.74	\pm 0.01$ \\
$K_1$ (km~s$^{-1}$)    	&$ 71.11 \pm 0.03 $    	&  \nodata 		        		& $71.11	\pm 0.03$ \\
$K_2$ (km~s$^{-1}$)    	&$ 71.90 \pm 0.03 $    	&  \nodata 		        		& $71.90	\pm 0.03$ \\
\enddata
\tablenotetext{*}{Fixed to spectroscopic solution.}
\end{deluxetable*}

We used this equation to model the squared visibilities and closure phases as a function of baseline and fit for the binary angular separation, position angle, and flux ratio for each observation using the grid search code\footnote{\href{http://chara.gsu.edu/analysis-software/binary-grid-search}{http://chara.gsu.edu/analysis-software/binary-grid-search}} of \citet{schaefer16}.   Based on the \textit{Hipparcos} parallax and the radii from \citet{fekel10}, the estimated angular diameters of both components are about $0.4$ mas. This is less than the $0.6$ mas resolution limit of CLIMB in the $K'$-band, so we set the angular diameters to be unresolved at $0$ mas. (We also tested finite angular diameters when fitting our data, but the results were consistent within the observational errors.)  The $u$ and $v$ coordinates are also known for each observation, so the only free parameters are the binary separations  and the flux ratio.   We first searched a wide range of relative separations, using MPFIT \citep{markwardt} to minimize the $\chi^2$ goodness-of-fit statistic in $V^2$ and CP at each point in the grid and find the best-fit $\Delta\alpha$,  $\Delta\delta$, and $f_2/f_1$.  We then calculated $\chi^2$ in a fine grid around this best-fit separation to determine the $1\sigma$ error ellipse from the positions where $\chi^2 \le \chi^2_{min} +1$.

 An example set of visibilities and closure phases are shown in Figure~\ref{vis} for the night of 2017 Sep 08. Our results for each night are listed in Table~\ref{relpos}, with the relative separation ($\rho$) and position angle ($\theta$) of the secondary component, the best-fit flux ratio ($f_2/f_1$), and the major axis ($\sigma_{max}$), minor axis ($\sigma_{min}$), and position angle ($\phi$) of the $1\sigma$ error ellipse. The position angles of the secondary component and error ellipse are both measured East of North. The weighted average flux ratio in $K'$-band is  $f_2/f_1 = 0.94 \pm 0.04$. On the nights of 2014 Oct 05 and 2017 Aug 04, only two brackets were observed, so the global $\chi^2$ map showed multiple solutions with $\chi^2 \le \chi^2_{min} +1$. In order to distinguish between these solutions, we predicted the relative separations from a preliminary orbit fit to the relative positions from nights with three or more brackets (see Section~\ref{vbfit}) and chose the solution closest to the predicted value.  Also, the flux ratio was not well constrained on 2017 Aug 04, so we held it fixed to the weighted average flux ratio.


 \begin{figure*}
\centering
\epsscale{1.0}
\plotone{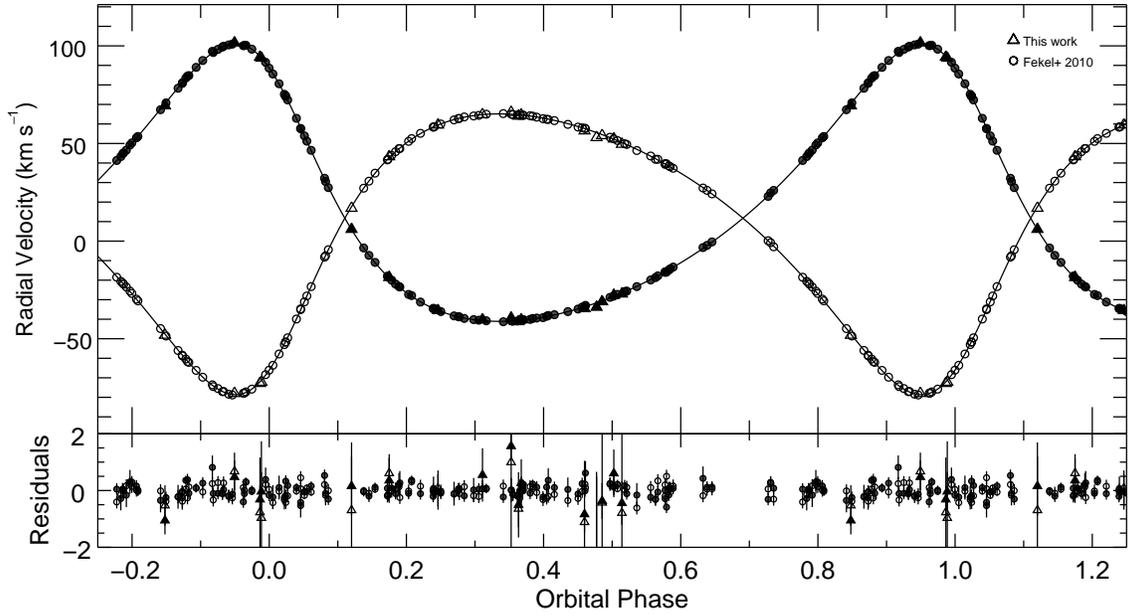}
\caption{Radial velocity curve for HD~224355 from the combined VB+SB2 solution. The filled and open points correspond to the observed velocities for the primary and secondary, where the triangles are the ARCES velocities and the circles are the velocities from \citet{fekel10}. The solid lines represent the model curves, and the residuals are shown in the bottom panel.  \label{rvcurve}} 
\vspace{0.5cm}
\end{figure*}

\begin{figure*}
\centering
\epsscale{1.0}
\plotone{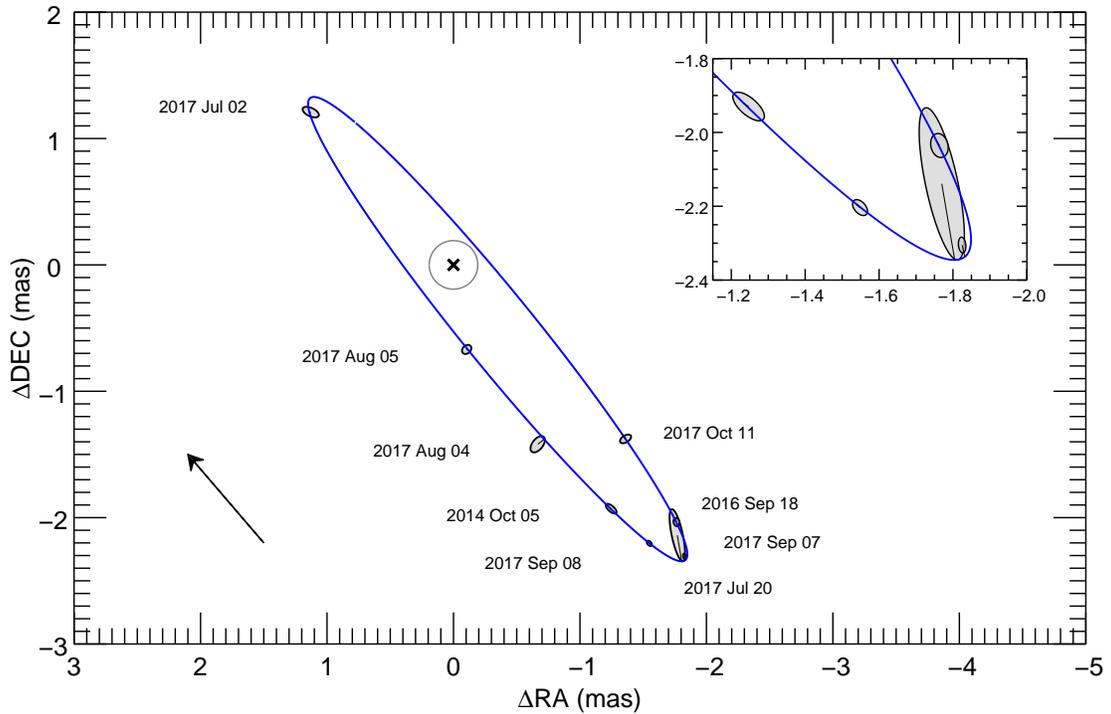}
\caption{Visual orbit for HD~224355 from the combined VB+SB2 solution. The primary star is located at the origin (black cross). The relative positions of the secondary are plotted as the filled ovals corresponding to the error ellipses, along with a line connecting the observed and predicted positions. The solid blue line shows the full model orbit, and the arrow shows the direction of orbital motion. The grey open circle around the origin represents the estimated angular size of the primary star.   The inset shows an expanded view of the lower portion of the orbit. \label{vb} }  
\vspace{0.8cm}
\end{figure*}

\section{Orbital Parameters}\label{section:orbit}
We first fit separately for the spectroscopic (SB2) and visual (VB) orbital parameters to ensure that our solutions were consistent with literature values, then performed a combined fit (VB+SB2) to determine the final orbital solution. Each step is explained below.

\subsection{Spectroscopic Orbit}\label{sbfit}
We fit for the spectroscopic orbital parameters of HD~224355 using the RVFIT code by \citet{rvfit}, which is an adaptive, simulated annealing code that fits for the parameters of single- and double-lined spectroscopic binaries\footnote{\href{http://www.cefca.es/people/~riglesias/rvfit.html}{http://www.cefca.es/people/$\sim$riglesias/rvfit.html}}.  We held the orbital period ($P$) fixed to the value from \citet{fekel10} and fit for the epoch of periastron ($T$), longitude of periastron of the primary ($\omega_1$), eccentricity ($e$), systemic velocity ($\gamma$), and velocity semi-amplitudes ($K_1$, $K_2$).  We then used the Monte Carlo Markov Chain (MCMC) feature of RVFIT to estimate the error in each parameter. The orbital elements determined from our ARCES radial velocities were consistent with those of \citet{fekel10}, making a joint solution possible.  The increased time baseline of a joint solution also allows for a more precise determination of the orbital period.

\citet{fekel10} did not give uncertainties for their radial velocities, but instead assigned weighting factors to the data from each instrument. We used $\sigma = 1/\sqrt{\rm weight}$ as first estimates of the uncertainties and ran RVFIT on their set of radial velocities.  We then rescaled the uncertainties such that $\chi^2_{red} = 1$, resulting in uncertainties of $0.2 - 0.4$ km~s$^{-1}$ which are reasonable for the high resolution of their spectra. 
We also added a correction of $-0.31$ km~s$^{-1}$ to the ARCES radial velocities so that the systemic velocity matched that of \citet{fekel10}. Finally, we fit for all of the spectroscopic orbital elements ($P, T, e, \omega_1, \gamma, K_1, K_2$) using RVFIT with the combined set of radial velocities. Our results are listed in the second column of Table~\ref{orbpar}, and are consistent with the results of \citet{fekel10} as expected.

\subsection{Visual Orbit}\label{vbfit}
We fit for the visual orbital elements using the procedure of \citet{schaefer16}, which uses the Newton-Raphson method to solve the equations of orbital motion and find the parameters that minimize $\chi^2$. We held the orbital period, eccentricity and longitude of periastron fixed to the spectroscopic solution and fit only for the orbital inclination ($i$), the angular semi-major axis ($a$), and the longitude of the ascending node ($\Omega$). Our results are listed in the third column of Table~\ref{orbpar}. We rescaled the uncertainties in relative position by a factor of 5.6 such that the reduced $\chi^2$ equals 1 in order to be used in the combined solution below. The parameter errors given in Table~\ref{relpos} are based upon these rescaled uncertainties.

\subsection{Combined VB + SB2 Solution}\label{vbsbfit}
Finally, we fit for all ten orbital parameters ($P$, $T$, $e$, $i$,  $a$, $\Omega$, $\omega_1$, $\gamma$, $K_1$, $K_2$) simultaneously using the Newton-Raphson method of \citet{schaefer16} to minimize $\chi^2$ in both the visual and spectroscopic orbits. We then performed a Monte Carlo error analysis, where we randomly varied each data point within its uncertainties (assuming Gaussian errors) and refit for the orbital parameters. We  created histograms of the best-fit parameters from several hundred thousand iterations, fit each histogram with a Gaussian, and took the standard deviation as the final $1\sigma$ uncertainty in each parameter. Our results are listed in the last column of Table~\ref{orbpar}. Figure~\ref{rvcurve} shows the best-fit radial velocity curve, and Figure~\ref{vb} shows the best-fit visual orbit.

\section{Derived Stellar Parameters}\label{section:param}

\subsection{Masses and Distance}	
Using the combined orbital solution of HD~224355, we derived stellar masses of $M_1 = 1.626 \pm 0.005 M_\odot$ and $M_2 = 1.608 \pm 0.005 M_\odot$ and a distance of $d=63.98 \pm 0.26$ pc.  Our distance from orbital parallax can be compared to the distances from trigonometric parallax in the literature; the distance is $71.0\pm1.8$ pc \citep{hipparcos2}  from \textit{Hipparcos} \citep{hipparcos1}, while the distance is $63.31^{+0.36}_{-0.35}$ pc \citep{bailer18} from \textit{GAIA} DR2 \citep{gaia1, gaia2}.

\begin{deluxetable}{lcc}[t!]
\tablewidth{0pt}
\tabletypesize{\normalsize}
\tablecaption{Stellar Parameters of HD 224355 \label{atmospar}}
\tablehead{ \colhead{Parameter} & \colhead{Primary} & \colhead{Secondary} }
\startdata		
Mass ($M_\odot$)			& $1.626 \pm 0.005$		& $1.608 \pm 0.005$ \\
Radius ($R_\odot$)			& $2.65\pm0.21$		& $2.47\pm0.23$	 \\
$\teff$ (K)					& $6450 \pm 120$ 		& $6590 \pm 110$ 	 \\
$\log g$ (cgs)				& $3.80\pm 0.04$		& $3.86 \pm 0.04 $ 	 \\ 
$V \sin i$ (km~s$^{-1}$)		& $ 10.9 \pm 1.2$		& $ 7.0 \pm 1.3$ 	 \\ 
Semi-major axis ($R_\odot$)  	&	 \multicolumn{2}{c}{$32.91 \pm 0.03$}   	 \\ 
Distance (pc)				&	 \multicolumn{2}{c}{$63.98 \pm 0.26$}	 \\ 
$E(B-V)$ (mag)				&	 \multicolumn{2}{c}{$0.04\pm0.05$}		 \\   
\enddata
\end{deluxetable}

\subsection{Radii and Surface Gravities}\label{sedfit}	
\begin{figure}
\centering
\epsscale{1.2}
\plotone{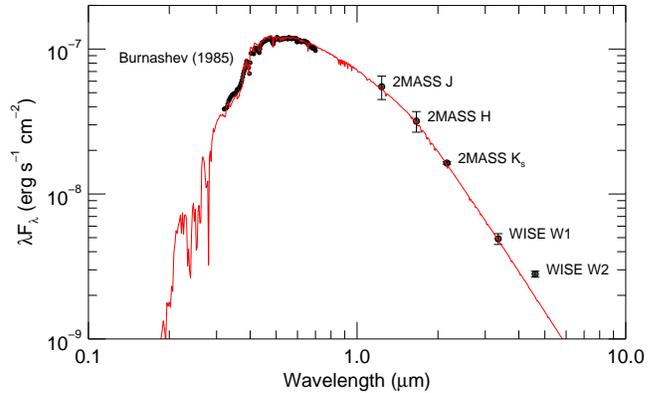}
\caption{Observed SED for HD 224355 (black points) and binary model (red line) using model atmospheres from \citet{sedmodel}. For clarity, the spectrophotometry error bars are not shown. \label{sed}} 
\end{figure}

In order to estimate the radius of each component, we used spectrophotometry and spectral energy distribution (SED) fitting. We combined optical spectrophotometry by \citet{burnashev85} with 2MASS \citep{2mass} and WISE \citep{wise} infrared magnitudes to create an SED of HD~224355.  Uncertainties of 5\% were adopted for the spectrophotometry.  The observed SED is shown as the black points in Figure~\ref{sed}. 

A model SED for a binary system is represented by 
$$f_\lambda = \frac{1}{d^2}  \Big(    R_1^2\ F_{\lambda 1} + R_2^2\ F_{\lambda 2} \Big) \times 10^{-0.4 A_\lambda}$$  where $F_{\lambda 1}$ and $F_{\lambda 2}$ are the surface fluxes of each component, $R_1$ and $R_2$ are the stellar radii, $d$ is the distance, and $A_\lambda$ is the extinction in magnitudes. The surface fluxes were taken from ATLAS9 model atmospheres \citep{sedmodel},  using the effective temperatures found in Section~\ref{tempfit} in an iterative process.  The radius ratio ($R_2/R_1$) can be calculated from the observed flux ratio and the model surface flux ratio. We calculated $R_2/R_1=0.94\pm0.06$ from the spectroscopic flux ratio (near H$\alpha$) and $R_2/R_1=0.90\pm0.10$ from the interferometric flux ratio (at 2.13$\mu m$), then found the weighted average to be $R_2/R_1 = 0.93\pm0.05$.  	

After substituting the average radius ratio into the equation above, we fit for the radius of the primary and the extinction using MPFIT.  Figure~\ref{sed} shows the best-fit binary SED model, and Table~\ref{atmospar} lists the best-fit parameters. We found stellar radii of $R_1 = 2.65\pm0.21 R_\odot$ and $R_2 = 2.47\pm0.23 R_\odot$, and surface gravities of $\log g_1 = 3.80\pm 0.04$ and $\log g_2 = 3.86 \pm 0.04$. The corresponding angular diameters of $\theta_1 = 0.38\pm0.03$ mas and $\theta_2 = 0.36\pm0.04$ mas are consistent with partial eclipses, as seen in the \textit{Hipparcos} light curve.  These radii are smaller than those found by \citet{fekel10} from colors and apparent magnitudes ($R_1 = 2.9 \pm 0.1 R_\odot$ and $R_2 = 2.8 \pm0.1 R_\odot$), likely because of the smaller parallax and lower temperatures used in their estimate. Additionally, we calculated the reddening to be $E(B-V)=0.04\pm0.05$ from the best-fit extinction and the Galactic extinction curve of \citet{redcurve}.

\subsection{Effective Temperatures and Rotational Velocities}\label{tempfit}
\begin{figure}
\centering
\epsscale{1.2}
\plotone{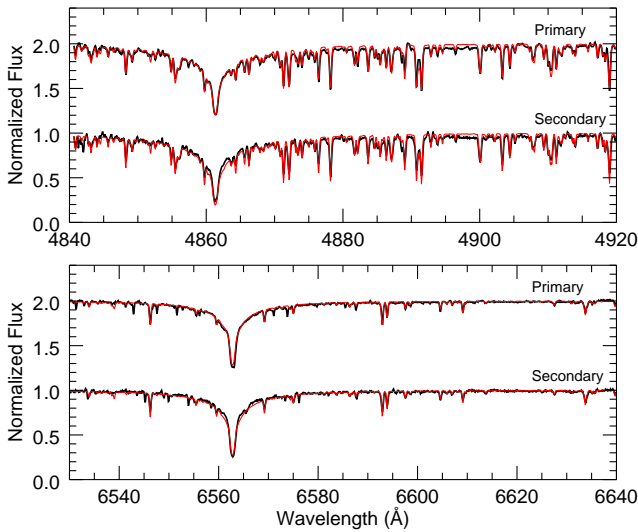}
\caption{Example reconstructed spectra of HD~224355 (black line) for the H$\beta$ (top) and H$\alpha$ (bottom) orders, as well as BLUERED model spectra (red line) with the atmospheric parameters listed in Table~\ref{atmospar}. \label{balmer}} 
\end{figure}

We reconstructed the spectrum of each component using the Doppler tomography algorithm of \citet{tomography} in order to determine the effective temperatures ($\teff$) and rotational velocities ($V \sin i$) of HD~224355. Template spectra were taken from BLUERED models using the atmospheric parameters in Table~\ref{atmospar} and solar metallicity. Example reconstructed spectra are shown in Figure~\ref{balmer}.

We first used line equivalent width ratios of several metal absorption lines to determine the effective temperatures of each component. We measured the equivalent widths ($W_\lambda$) of these lines in both the reconstructed spectra and model spectra of different effective temperatures using the ARES code\footnote{\href{http://www.astro.up.pt/~sousasag/ares/}{http://www.astro.up.pt/$\sim$sousasag/ares/}} of \citet{ares}. For each pair of absorption lines, we calculated the $W_\lambda$ ratio as a function of effective temperature and interpolated between model ratios to determine the effective temperatures of each component. Each pair was also weighted between $0-1$ based on how fast the ratio changed with temperature, such that line pairs more sensitive to temperature have higher weights. These weights were then used to calculate the weighted mean effective temperature for each component and the uncertainties corresponding to the standard deviation of the results from all line ratios.

\begin{figure*}
\centering
\epsscale{1.1}
\plotone{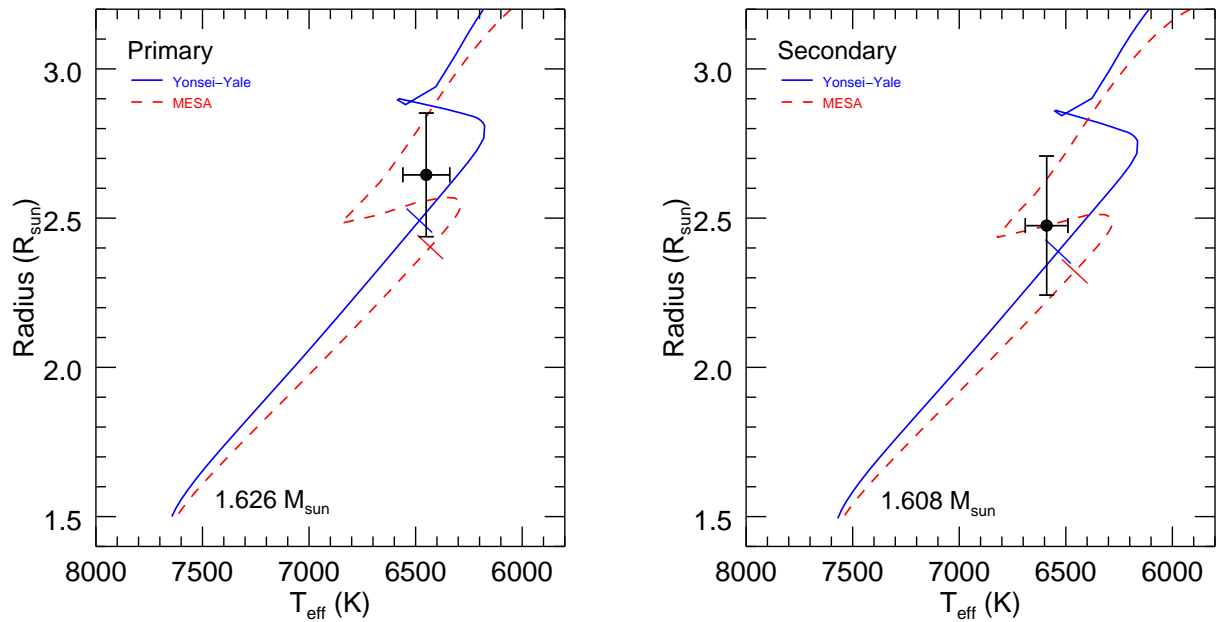}
\caption{Evolutionary tracks for the primary component (left) and secondary component (right) of HD~224355. The Yonsei-Yale $Y^2$ models are the blue solid lines and the MESA models are the red dashed lines.    The tick marks correspond to the mean age of the binary for each model. \label{evo}}
\vspace{0.5cm}
\end{figure*}

Next, we determined the projected rotational velocities of each component by fitting model spectra of various $V \sin i$ to the reconstructed spectra. We chose 50 metal absorption lines in the red part of the spectrum that are not blended and have a well-defined continuum, mostly \ion{Fe}{1} or \ion{Fe}{2}.  For each line, we calculated $\chi^2$ of each model as a function of $V \sin i$, then fit a parabola to the curve to determine the  $V \sin i$ corresponding to the minimum $\chi^2$ and the uncertainty corresponding to $\chi_{min}^2+1$. We found the weighted average $V \sin i$ of each component to be $V_1 \sin i = 10.9 \pm 1.2$ km~s$^{-1}$ and $V_2 \sin i = 7.0 \pm 1.3$ km~s$^{-1}$.  The primary component of HD~224355 is rotating at the projected synchronous velocity of 10.9  km~s$^{-1}$ and the secondary is rotating slower than the projected synchronous velocity of 10.2 km~s$^{-1}$, which is consistent with the trend in rotational velocities of \textit{Kepler} binaries \citep{lurie17}.

\subsection{Comparison with Evolutionary Models} 

We compared the observed parameters of HD 224355 to both the Yonsei-Yale $Y^2$ \citep{y2} and MESA \citep{mesa1, mesa4} stellar evolution codes.  The Yonsei-Yale models\footnote{\href{http://www.astro.yale.edu/demarque/yystar.html}{http://www.astro.yale.edu/demarque/yystar.html}} were created using their model interpolator, shown as the solid lines in Figure~\ref{evo}.  These models use the step-function method to characterize convective core overshooting as a function of mass and metallicity, where $\Lambda_{\rm ov}=0.2$ for both components. The MESA models\footnote{\href{http://www.mesa.sourceforge.net}{http://www.mesa.sourceforge.net}} were computed at the observed masses and shown as the dashed lines in Figure~\ref{evo}. MESA uses the diffusion method to characterize convective core overshooting, so we estimated the overshooting parameter of both components to be $f_{ov}= 0.01$ from the calibration of \citet{ct18}. Both sets of models are non-rotating and use solar metallicity and scaled solar abundances.

We estimated the age of each component based on which points lie within the observed uncertainties, then calculated the mean age of the system for each set of models (noted as the tick marks in Figure~\ref{evo}). For the Yonsei-Yale models, both components of HD 224355 appear to lie towards the end of the main sequence.  The individual component ages are $1.92$ and $1.86$ Gyr with a mean system age of $1.89$ Gyr. For the MESA models, the components intersect the evolutionary tracks at the end of the main sequence and twice on the blue hook. We chose the main sequence solution because it yields the closest ages between the components. We found individual ages of $1.64$ and $1.51$ Gyr with a mean system age of $1.58$ Gyr.

\section{Discussion}
We determined the mass of each component to within $0.3\%$ error and the radius of each component to within $9\%$ error by combining the visual orbit from CHARA observations with the spectroscopic orbit. While the uncertainties in mass are sufficiently small, the uncertainties in radius are not small enough for a critical test of stellar evolution models. Future interferometric observations in the optical could more precisely measure the stellar radii; for example, the PAVO beam combiner at CHARA has an angular resolution of $0.2$ mas and would be able to resolve both components. The component radii could also be found from light curve modeling. The \textit{Hipparcos} photometry did not cover the eclipse of the secondary component, so we encourage observers to obtain more photometry for this system to expand the phase coverage and allow for eclipse modeling. 

A possible source of error in our analysis would be the presence of an unknown tertiary companion. Flux from a third component would dampen the interferometric fringe visibilities, bias the measured flux ratio, and add absorption features to the spectra. Furthermore, unaccounted flux might lead to overestimates of the radii
derived from the SED fit (Section~\ref{sedfit}). We do not see any evidence of a third component in our spectra, but upcoming observations using the 'Alopeke speckle camera on Gemini North will confirm or rule out the presence of a tertiary companion. 'Alopeke can resolve companions down to 16 mas, in which case the effects would be seen in the CLIMB observations in the form of separated fringe packets.

Our results demonstrate the value of studies of resolved systems for our goal of comparing the fundamental parameters of short and long period binaries by measuring the visual orbits of spectroscopic binaries. These visual orbits also provide model-independent distances from orbital parallax which can be compared to trigonometric and spectroscopic parallaxes.  For this purpose, we are continuing observations at CHARA and APO to resolve the visual and spectroscopic orbits of several other bright binary systems to determine their fundamental parameters.
\smallskip

\acknowledgments
{\small

The authors would like to thank the staff at APO and CHARA for their help during observations, and we are grateful to an anonymous referee for their valuable comments. This work is based upon observations obtained with the Georgia State University Center for High Angular Resolution Astronomy Array at Mount Wilson Observatory. The CHARA Array is supported by the National Science Foundation under Grants No. AST-1636624 and AST-1715788. Institutional support has been provided from the GSU College of Arts and Sciences and the GSU Office of the Vice President for Research and Economic Development. This work has made use of the Jean-Marie Mariotti Center SearchCal service, the CDS Astronomical Databases SIMBAD and VIZIER, data from the Wide-field Infrared Survey Explorer, and data from the Two Micron All Sky Survey. }

\facilities{CHARA, APO:3.5m}

\software{ 
ARES \citep{ares},  
Grid Search for Binary Stars \citep{schaefer16},   
MESA \citep{mesa1, mesa4},  
MPFIT \citep{markwardt},  
RVFIT \citep{rvfit},  
SearchCal \citep{searchcal}, 
TODCOR \citep{todcor}, 
Y$^2$ models \citep{y2} 
}



\begin{thebibliography}{}

\bibitem[Bagnuolo et al.(1992)]{tomography} 
Bagnuolo, W.~G., Jr., Gies, D.~R., \& Wiggs, M.~S.\ 1992, \apj, 385, 708 

\bibitem[Bailer-Jones et al.(2018)]{bailer18}
Bailer-Jones, C.~A.~L., Rybizki, J., Fouesneau, M., Mantelet, G., \& Andrae, R.\ 2018, \aj, 156, 58 

\bibitem[Bertone et al.(2008)]{bluered} 
Bertone, E., Buzzoni, A., Ch{\'a}vez, M., \& Rodr{\'{\i}}guez-Merino, L.~H.\ 2008, \aap, 485, 823  

\bibitem[Boden et al.(1999)]{boden99} 
Boden, A.~F., Lane, B.~F., Creech-Eakman, M.~J., et al.\ 1999, \apj, 527, 360 

\bibitem[Boden(2000)]{boden00} 
Boden, A.~F.\ 2000, Principles of Long Baseline Stellar Interferometry, ed. P. R. Lawson (Pasedena: NASA/JPL and CalTech), 9

\bibitem[Burnashev(1985)]{burnashev85} 
Burnashev, V.~I.\ 1985, Abastumanskaia Astrofizicheskaia Observatoriia Byulleten, 59, 83 

\bibitem[Castelli \& Kurucz(2004)]{sedmodel} 
Castelli, F., \& Kurucz, R.~L.\ 2004, arXiv:astro-ph/0405087 

\bibitem[Chelli et al.(2016)]{searchcal} 
Chelli, A., Duvert, G., Bourg{\`e}s, L., et al.\ 2016, \aap, 589, A112 

\bibitem[Chaplin \& Miglio(2013)]{astero} 
Chaplin, W.~J., \& Miglio, A.\ 2013, \araa, 51, 353 

\bibitem[Claret \& Torres(2016)]{ct16} 
Claret, A., \& Torres, G.\ 2016, \aap, 592, A15 

\bibitem[Claret \& Torres(2018)]{ct18} 
Claret, A., \& Torres, G.\ 2018, \apj, 859, 100 

\bibitem[Demarque et al.(2004)]{y2} 
Demarque, P., Woo, J.-H., Kim, Y.-C., \& Yi, S.~K.\ 2004, \apjs, 155, 667 

\bibitem[Eker et al.(2015)]{eker15} 
Eker, Z., Soydugan, F., Soydugan, E., et al.\ 2015, \aj, 149, 131 

\bibitem[Enoch et al.(2010)]{enoch10} 
Enoch, B., Collier Cameron, A., Parley, N.~R., \& Hebb, L.\ 2010, \aap, 516, A33 

\bibitem[Fekel et al.(2010)]{fekel10} 
Fekel, F.~C., Tomkin, J., \& Williamson, M.~H.\ 2010, \aj, 139, 1579 

\bibitem[Fitzpatrick(1998)]{redcurve} 
Fitzpatrick, E.~L.\ 1998, Ultraviolet Astrophysics Beyond the IUE Final Archive, 413, 461 

\bibitem[Gaia Collaboration et al.(2016)]{gaia1} 
Gaia Collaboration, Prusti, T., de Bruijne, J.~H.~J., et al.\ 2016, \aap, 595, A1 

\bibitem[Gaia Collaboration et al.(2018)]{gaia2} 
Gaia Collaboration, Brown, A.~G.~A., Vallenari, A., et al.\ 2018, \aap, 616, A1 

\bibitem[Halbwachs(1981)]{halbwachs81} 
Halbwachs, J.~L.\ 1981, \aaps, 44, 47 

\bibitem[Harper(1923)]{harper23} Harper, W.\ 1923, Publications of the Dominion Astrophysical Observatory Victoria, 2, 263 

\bibitem[Hilditch(2001)]{hilditch} Hilditch, R.~W.\ 2001, \href{http://adsabs.harvard.edu/abs/2001icbs.book.....H}{An Introduction to Close Binary Stars} (Cambridge, UK: Cambridge University Press)

\bibitem[Hummel et al.(1993)]{hummel93} 
Hummel, C.~A., Armstrong, J.~T., Quirrenbach, A., et al.\ 1993, \aj, 106, 2486 

\bibitem[Hurley et al.(2002)]{hurley02} 
Hurley, J.~R., Tout, C.~A., \& Pols, O.~R.\ 2002, \mnras, 329, 897 

\bibitem[Iglesias-Marzoa et al.(2015)]{rvfit} 
Iglesias-Marzoa, R., L{\'o}pez-Morales, M., \& Jes{\'u}s Ar{\'e}valo Morales, M.\ 2015, \pasp, 127, 567 

\bibitem[Imbert(1977)]{imbert77} Imbert, M.\ 1977, \aaps, 29, 407 

\bibitem[Kolbas et al.(2015)]{kolbas15} 
Kolbas, V., Pavlovski, K., Southworth, J., et al.\ 2015, \mnras, 451, 4150 

\bibitem[Lurie et al.(2017)]{lurie17} 
Lurie, J.~C., Vyhmeister, K., Hawley, S.~L., et al.\ 2017, \aj, 154, 250 

\bibitem[Markwardt(2009)]{markwardt} Markwardt, C.~B.\ 2009, in ASP Conf. Ser. 411, Astronomical Data Analysis Software and Systems XVIII, ed. D. A. Bohlender, D. Durand, \& P. Dowler (San Francisco, CA: ASP), 251

\bibitem[Monnier et al.(2011)]{monnier11} 
Monnier, J.~D., Zhao, M., Pedretti, E., et al.\ 2011, \apjl, 742, L1 

\bibitem[Kluska et al.(2018)]{kluska18} 
Kluska, J., Kraus, S., Davies, C.~L., et al.\ 2018, \apj, 855, 44 

\bibitem[Moya et al.(2018)]{moya18} 
Moya, A., Zuccarino, F., Chaplin, W.~J., \& Davies, G.~R.\ 2018, \apjs, 237, 21 

\bibitem[Otero (2006)]{otero06} Otero, S. 2006, IBVS, 5699, 1

\bibitem[Paxton et al.(2011)]{mesa1} 
Paxton, B., Bildsten, L., Dotter, A., et al.\ 2011, \apjs, 192, 3 

\bibitem[Paxton et al.(2018)]{mesa4} 
Paxton, B., Schwab, J., Bauer, E.~B., et al.\ 2018, \apjs, 234, 34 

\bibitem[Perryman et al.(1997)]{hipparcos1} 
Perryman, M.~A.~C., Lindegren, L., Kovalevsky, J., et al.\ 1997, \aap, 323, L49

\bibitem[Plaskett et al.(1920)]{plaskett20} 
Plaskett, J.~S., Harper, W.~E., Young, R.~K., \& Plaskett, H.~H.\ 1920, Publications of the Dominion Astrophysical Observatory Victoria, 1, 163 

\bibitem[Raghavan et al.(2009)]{deepak09} 
Raghavan, D., McAlister, H.~A., Torres, G., et al.\ 2009, \apj, 690, 394 

\bibitem[Torres et al.(2010)]{torres10} 
Torres, G., Andersen, J., \& Gim{\'e}nez, A.\ 2010, \aapr, 18, 67 

\bibitem[Schaefer et al.(2016)]{schaefer16} 
Schaefer, G.~H., Hummel, C.~A., Gies, D.~R., et al.\ 2016, \aj, 152, 213 

\bibitem[Skrutskie et al.(2006)]{2mass} 
Skrutskie, M.~F., Cutri, R.~M., Stiening, R., et al.\ 2006, \aj, 131, 1163 

\bibitem[Sousa et al.(2007)]{ares} 
Sousa, S.~G., Santos, N.~C., Israelian, G., Mayor, M., \& Monteiro, M.~J.~P.~F.~G.\ 2007, \aap, 469, 783 

\bibitem[ten Brummelaar et al.(2005)]{chara} 
ten Brummelaar, T.~A., McAlister, H.~A., Ridgway, S.~T., et al.\ 2005, \apj, 628, 453 

\bibitem[ten Brummelaar et al.(2013)]{climb} 
ten Brummelaar, T.~A., Sturmann, J., Ridgway, S.~T., et al.\ 2013, 
Journal of Astronomical Instrumentation, 2, 1340004 

\bibitem[Tody(1986)]{iraf1}
Tody, D.\ 1986, in Instrumentation in astronomy VI, Proc. SPIE 0627, ed. D. L. Crawford (Bellingham, WA: SPIE), 733

\bibitem[Tody(1993)]{iraf2} 
Tody, D.\ 1993, in Astronomical Data Analysis Software and Systems II, ASP Conf. Vol. 52, ed. R. J. Hanisch, R. J. V. Brissenden, \& J. Barnes (San Francisco: ASP), 173

\bibitem[Tokovinin et al.(2006)]{tokovinin06} 
Tokovinin, A., Thomas, S., Sterzik, M., \& Udry, S.\ 2006, \aap, 450, 681 

\bibitem[Torres et al.(2010)]{torres10} 
Torres, G., Andersen, J., \& Gim{\'e}nez, A.\ 2010, \aapr, 18, 67 

\bibitem[van Leeuwen(2007)]{hipparcos2} 
van Leeuwen, F.\ 2007, \aap, 474, 653 

\bibitem[Wang et al.(2003)]{arces} 
Wang, S.-i., Hildebrand, R.~H., Hobbs, L.~M., et al.\ 2003, in Instrument Design and Performance for Optical/Infrared Ground-based Telescopes, Proc. SPIE 4841, ed. M. Iye \& A. F. M. Moorwood (Bellingham, WA: SPIE), 1145

 \bibitem[Wright et al.(2010)]{wise} 
 Wright, E.~L., Eisenhardt, P.~R.~M., Mainzer, A.~K., et al.\ 2010, \aj, 140, 1868 

\bibitem[Zucker \& Mazeh(1994)]{todcor} 
Zucker, S., \& Mazeh, T.\ 1994, \apj, 420, 806 

\end{thebibliography}
\end{document}